# HIGHER ORDER MODES FOR BEAM DIAGNOSTICS IN THIRD HARMONIC 3.9 GHZ ACCELERATING MODULES[*]

N. Baboi[#], N. Eddy[§], T. Flisgen[°], H.-W. Glock[°], R. M. Jones[†‡], I. R. R. Shinton[†‡], and P. Zhang[†#‡]

[#] Deutsches Elektronen-Synchrotron (DESY), Hamburg, Germany
[§] Fermi National Accelerator Laboratory (Fermilab), Batavia, Illinois, U.S.A.
[°] Universität Rostock, Rostock, Germany
[†] School of Physics and Astronomy, The University of Manchester, Manchester, U.K.
[‡] The Cockcroft Institute of Accelerator Science and Technology, Daresbury, U.K.

## Abstract

An international team is currently investigating the best way to use Higher Order Modes (HOM) for beam diagnostics in 3.9 GHz cavities. HOMs are excited by charged particles when passing through an accelerating structure. Third harmonic cavities working at 3.9 GHz have been installed in FLASH to linearize the bunch energy profile. A proof-of-principle of using HOMs for beam monitoring has been made at FLASH in the TESLA 1.3 GHz cavities. Since the wakefields generated in the 3.9 GHz cavities are significantly larger, their impact on the beam should be carefully minimized. Therefore our target is to monitor HOMs and minimize them by aligning the beam on the cavity axis. The difficulty is that, in comparison to the 1.3 GHz cavities, the HOM-spectrum is dense, making it difficult to identify individual modes. Also, most modes propagate through the whole cryo-module containing several cavities, making it difficult to measure local beam properties. In this paper the options for the HOM-based beam position monitors are discussed.

## INTRODUCTION

Higher Order Modes (HOM) are electromagnetic eigenmodes excited by charged particle bunches passing through accelerating structures. They can damage the beam quality, but can also be used to measure beam properties, for example its transverse position.

This paper presents the concept and options of HOM-based beam position monitors (HOM-BPMs) for the superconducting 3.9 GHz cavities at the Free-electron LASer in Hamburg (FLASH).

## FLASH

FLASH is a self-amplified spontaneous emission free electron laser delivering ultra-short laser-like pulses with a wavelength between about 4.5 and 47 nm [1,2]. It is a user and a test facility. Fig. 1 shows the main components of the linac. A photoelectric gun generates trains of electron bunches. The charge of each bunch is between 100 pC and 1 nC. The pulses are up to 800 μs long and have a repetition frequency of 10 Hz. The bunch repetition frequency within the pulses is up to 1 MHz.

The electrons are accelerated to up to 1200 MeV by 7 accelerating structures (cryo-modules), each containing eight super-conducting 1.3 GHz TESLA cavities. Two bunch compressors reduce the length of each bunch to the order of hundreds of femtoseconds. Four 3rd harmonic cavities working at 3.9 GHz are used to linearize the longitudinal phase space. The beam passes through transverse and energy collimators and then produces laser pulses in special undulators. The photons continue to the FEL experimental hall, while the electron beam is sent to the dump. A bypass is used to protect the undulators during special experiments. sFLASH is a seeded experiment.

## 3rd Harmonic Cavities

Just behind the first accelerating module, cryo-module ACC39 containing the four 3rd harmonic superconducting cavities is located. These are shown schematically in Fig. 2. Each 9-cell cavity has at one side a power coupler and at each side a HOM coupler, extracting power from the beam-excited fields, other than the fundamental mode at 3.9 GHz.

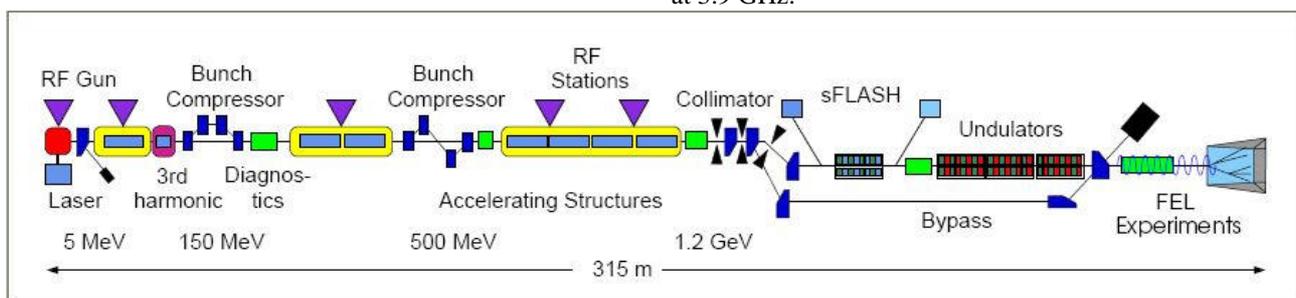

Figure 1: Layout of the FLASH linac [2].

___________________________________________
*Work supported in part by the European Commission within the Framework Programme 7, Grant Agreement 227579.

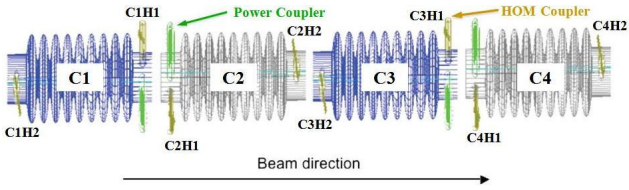

Figure 2: Schematic representation of the four $3^{rd}$ harmonic cavities in ACC39. The power couplers (green) are placed downstream for cavities 1 and 3 (C1 and C3) and upstream for C2 and C4. The HOM couplers (brown) on the same side with the power couplers are named H1, the other ones H2.

Figure 3 shows a picture of a 3.9 GHz cavity in comparison to a TESLA 1.3 GHz cavity. The $3^{rd}$ harmonic cavities designed at Fermilab in collaboration with DESY [3,4] were scaled down by a factor of 3 with respect to the TESLA cavities, so that the fundamental frequency is 3.9 GHz. Two types of HOM couplers have been designed and installed: one for C1 and C3 and the other for C2 and C4 [5].

The wakefields of the $3^{rd}$ harmonic cavities are expected to be larger than those of the TESLA cavities [6]. Therefore special care has been paid to designing the HOM-couplers. Also, the beam pipes at either side of the cavities are more than a factor 3 of those of the TESLA cavities. In this way most HOMs are above the cut-off frequencies of the beam pipe and are able to reach the HOM couplers of other cavities, which also dump them. This and the careful cavity design have led to lower quality factors (typically of the order of $10^4$) than for the TESLA cavities (typically $10^5$). While this is good for the beam dynamics, it has consequences on the design of the HOM-BPMs, as discussed later in this paper.

## DIPOLE MODES FOR THE HOM-BPMS AT THE 3.9 GHZ CAVITIES

Many theoretical and experimental investigations have been made on the spectrum of individual $3^{rd}$ harmonic cavities and of the entire module ACC39 [9-12]. Fig. 4 shows one beam excited spectrum as measured from C4H1. The eigenmodes of the monopole, dipole and quadrupole modes for an ideal single cavity are also marked.

Most interesting for use as beam position monitors (BPMs) are dipole modes, whose amplitude is proportional to the exciting beam offset and charge [7,8]. The principle is similar to that employed in a cavity BPM, where a typically well separated dipole mode with two polarizations, one horizontal and one vertical, is used. In the 3.9 GHz cavities the dipole mode amplitude is also proportional to the beam offset. Fig. 5 illustrates this for a mode at 4.73 GHz. However in a multi-cell accelerating cavity the polarizations are rarely horizontal and vertical and the spectrum is rather complex. Therefore understanding the HOM spectrum is important.

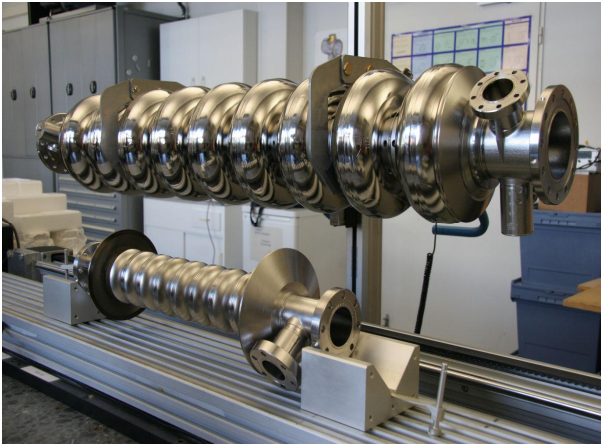

Figure 3: TESLA cavity (upper) and $3^{rd}$ harmonic cavity (lower). On the right side of each cavity one can see a port for the input coupler. The HOM couplers can also be partially seen at either side of the cavities.

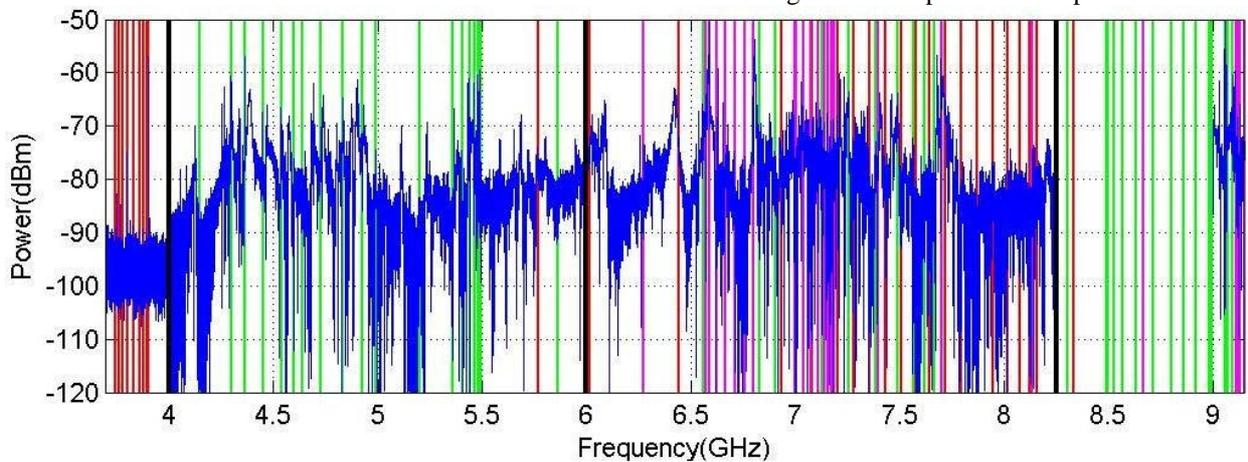

Figure 4: Beam-excited spectrum from C4H1. The vertical lines mark the theoretical frequencies of the monopole (red), dipole (green) and quadrupole (magenta) modes for an ideal single cavity without couplers. The black lines divide the spectrum into 4 parts, which are measured under different beam charge and offset or signal attenuation. No data have been taken between about 8.25 and 9 GHz.

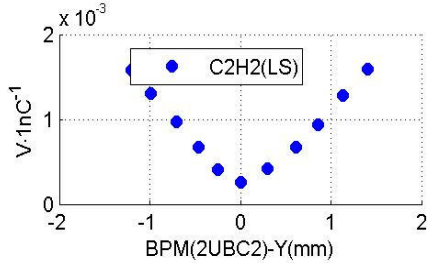

Figure 5: Amplitude of a cavity dipole mode at 4.73 GHz, normalized to the charge, as measured from C2H2, as a function of the beam offset, as measured by the downstream BPM (2UBC2).

*Cavity Modes*

The first two dipole bands are of highest interest since they contain the modes with highest R/Q (see Table 1). These couple strongest to the beam, which gives the potential for higher resolution of the beam position monitoring.

Table 1: Dipole modes with highest R/Q in the first two cavity dipole passbands (simulations for single cavity without couplers).

| Dipole passband | Frequency [GHz] | R/Q [$\Omega/cm^2$] |
| --- | --- | --- |
| 1st band | 4.723 | 10.37 |
| | 4.831 | 50.20 |
| | 4.926 | 30.38 |
| 2nd band | 5.444 | 20.88 |
| | 5.470 | 16.07 |

Previously, based on the same idea, HOM-BPM electronics has been built for the TESLA cavities at FLASH by SLAC [13]. There, a high R/Q mode (R/Q = 5.5 $\Omega/cm^2$ [14]) at about 1.7 GHz from the first dipole band has been filtered out of the spectrum and used for beam monitoring. The principle has been successfully demonstrated. However translating the concept to the 3.9 GHz cavities proved difficult. In order to explain why, we first show in Fig. 6 the beam excited spectrum of the first dipole band of a 3rd harmonic cavity (C4H1, Fig. 6a) and of a TESLA cavity (Fig. 6b). For the 1.3 GHz cavity one can easily identify the modes in the passband. When zooming in around each peak, one can usually distinguish both polarizations. For different cavities there is a spread in the mode frequencies. The electronics for the TESLA cavities has a bandwidth of 20 MHz, which includes the high R/Q mode for each cavity.

For the 3rd harmonic cavity the modes are not well separated. This is due to several reasons: The individual modes are overlapping, due to the higher frequency and to the lower Q factors (typically of the order of $10^4$, while for the TESLA cavities they are of the order of $10^5$). The modes are confined inside the cavity for the TESLA structures, while for the 3.9 GHz ones the modes propagate between the cavities, due to the larger beam pipe diameter than would be the case if scaling down the beam pipe of the TESLA cavity.

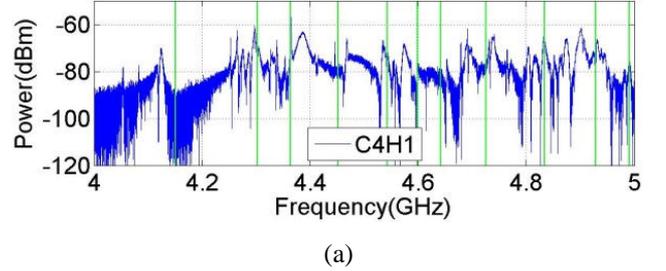

(a)

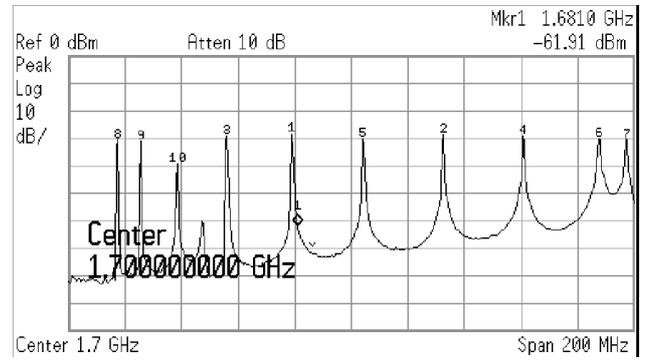

(b)

Figure 6: Beam-excited spectrum for the first dipole passband of the 3rd harmonic cavity C4H1 (a – detail from Fig. 4) and of a TESLA cavity (b).

Figure 7 shows a comparison of the measured $S_{21}$ scattering parameter for the first dipole band across a single 3rd harmonic cavity, as measured in the single cavity test facility at Fermilab and in the module at the Cryo-Module Test Bench (CMTB) at DESY [9]. One notices that more peaks are present in the module-based measurement. For the isolated cavity case, some modes are separated from each other, while for the module-based measurement the spectrum is more continuous. This is due to the fact that most modes propagate through the beam pipes, whose theoretical cut-off frequency is 4.39 GHz.

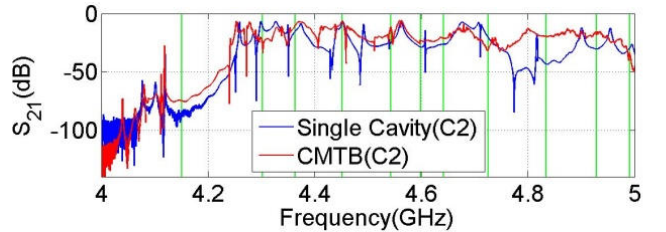

Figure 7: $S_{21}$ scattering parameter measured between C2H1 and C2H2 during the single cavity tests at Fermilab and in the cryo-module at CMTB at DESY.

In practice the cut-off frequency is even lower, due to the presence of couplers. Basically all modes of the first dipole passband propagate. This is seen in transmission measurements across the full module (from C1H2 to C4H2) and in simulations [11,15].

The complicated spectrum has several consequences for the HOM-BPM electronics to be built: One cannot separate and identify with certitude the high R/Q modes; and the localized measurement of the beam position is rather difficult.

Therefore we have the option to select the frequency region around the theoretical frequency with high R/Q, and target a potential good resolution. Or use localized modes which could allow for a local beam position measurement, at the expense of resolution. Several such modes are listed in Table 2.

Table 2: Localized dipole modes which could be used for beam position monitoring (simulations for single cavity without couplers).

| Mode type | Frequency [GHz] | R/Q [$\Omega/cm^2$] |
|---|---|---|
| Beam pipe | 4.1486 | 0.24 |
|  | 4.1487 | 1.31 |
| Cavity / 5th band | 9.057 | 0.05 |
|  | 9.059 | 0.07 |
|  | 9.062 | 2.17 |
|  | 9.070 | 4.04 |

*Beam Pipe Modes*

One option to get a localized measurement of the beam position is to use beam pipe dipole modes. These are localized in the beam pipes between the cavities. Such modes can be seen in Fig. 6a below 4.2 GHz. Simulations show that they have a rather low R/Q factor (see Table 2). These modes have shown also a good linearity with the beam offset [12].

*Trapped Cavity Modes*

In spite of being above the cut-off of the beam pipe, some modes in the 5th dipole band are also localized (see Table 2). This is due to the fact that there is little field in the end cells. Some signal reaches however the HOM couplers, so that they can be monitored. Such modes showed a good correlation to the beam position also [16].

## HOM-BPMS FOR ACC39

As stated in the previous section, there are three options for the HOM-BPMs for ACC39:

- High R/Q propagating cavity dipole modes from the 1st or 2nd dipole band: They have the potential of high resolution of the beam position measurement, but the measurement is not local, since the fields propagate in the whole module.
- Beam pipe modes: These have the advantage of enabling a local measurement of the beam position, but give a lower resolution and are not cavity based, therefore less relevant for the impact of cavity modes on the beam.
- Trapped cavity modes of the 5th dipole band: These also make a local beam position measurement possible and are cavity based, but should give lower resolution.

In all cases, we would most likely not separate one single mode, but a group of modes. Alternative electronics concepts are investigated.

*Mode Selective Electronics*

A system based upon the narrowband downmix electronics developed for the FLASH TESLA cavities is being developed [13]. The new system is being designed with the flexibility to examine multiple modes of interest as well as accommodate the large mode bandwidths observed in the 3.9 GHz cavities. This has led to a design with four switchable analog filter sections (see Fig. 8).

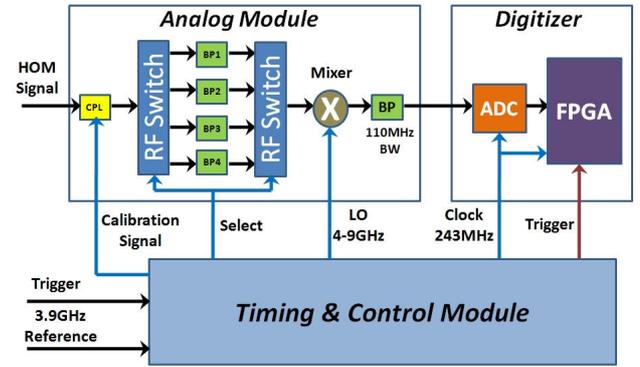

Figure 8: Block diagram of proposed mode selective narrowband electronics for the 3.9 GHz cavities.

The system consists of three basic components: an analog module to select analog bandpass filters and mix the signal to an intermediate frequency of less than 125 MHz; a 14 bit digitizer operating at up to 250 MS/s with a programmable FPGA to allow for additional digital filtering and signal processing; and a timing and control module to generate selectable LO and calibration signals from 4-9 GHz along with a digitizer clock of 243 MHz locked to the 3.9 GHz cavity RF and synchronized trigger signals.

Currently the design and layout for the analog and timing modules are under study. To provide the flexibility to select modes from 4-9 GHz is challenging as there are a limited number of components able to operate over this whole frequency range. Also, the bandwidths required (up to 110 MHz) make isolating the HOM modes to achieve high resolution challenging as well as it is necessary to filter out one mixer side-band.

Ideally, one would also like to recover the phase as well as magnitude information for the modes. This necessitates having the mixer LO locked to the machine RF with a locked calibration signal to monitor phase drifts due to cables and electronics. This is a critical requirement for the 1.3 GHz TESLA cavities with clearly identified modes, but may be very difficult, if not impossible, to measure for 3.9 GHz cavity modes which propagate throughout the module.

*Diode-based Signal Capturing*

In addition to the mode-selective approach described above, a second signal capturing and evaluation scheme is under investigation [17]. Here only a broadband filter and a RF-detector diode (AGILENT 423B) together with a conventional digitizing oscilloscope are used to analyze the HOM coupler signals (compare Fig. 9). The diode's output voltage is proportional to the time-dependent total HOM power, being a complicated superposition of all modes within the bandwidth of the entrance filter. The only purpose of this filter is the suppression of the 3.9 GHz fundamental mode signal, which is dominant in spite of the HOM coupler's own notch filter.

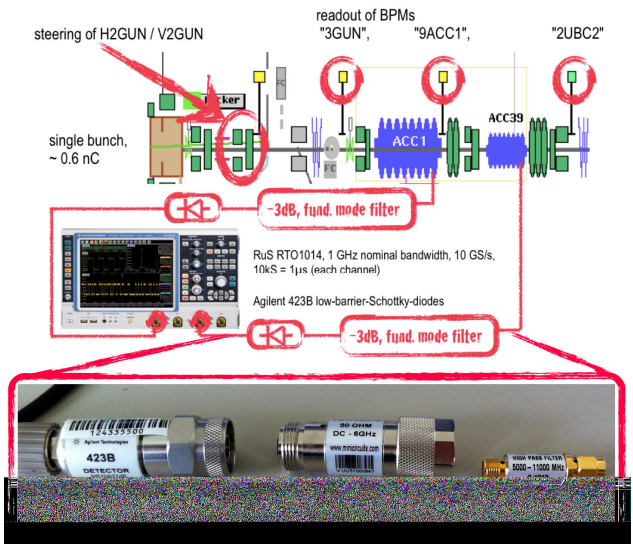

Figure 9: Experimental setup used for the diode detection. The acquisition is triggered by the HOM signal.

The price to be paid for this least-effort setup is the analysis of a complicated signal shape, which also could be interpreted as a sum of all possible mixing products in the HOM spectrum. Therefore the dominant spectral content of the diode output signal lies in the low-frequency range below some hundreds MHz.

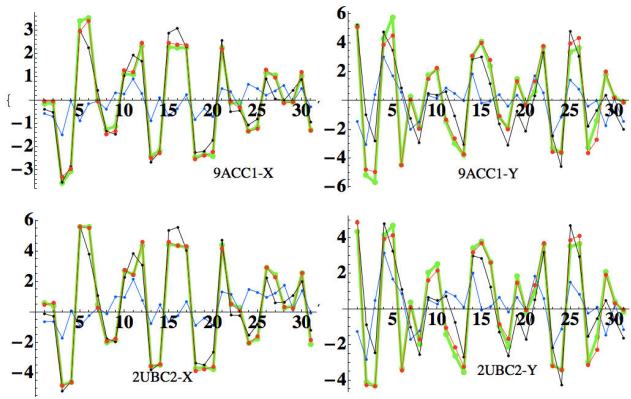

Figure 10: BPM readout (green: upstream 9ACC1, downstream 2UBC2) compared with linearly correlated HOM-SVD-based prediction (from C4H2) using 5 (blue), 7 (black) and 10 (red) singular values (mm vs. sample).

It was recently shown that a Singular-Value-Decomposition (SVD) of this time dependence is very well described as a superposition of only a few (roughly 10) dominant signal components [17]. Their amplitudes were demonstrated to be quite linearly correlated to BPM readouts in both transverse coordinates up- and downstream of ACC39 (compare Fig. 10). This means that a single HOM coupler port allows for the observation of both transverse beam position and charge.

## SUMMARY AND OUTLOOK

A series of studies with and without beam on dipole modes in ACC39 has been performed. Three kinds of dipole modes are considered for use as HOM-BPMs: propagating cavity modes, beam pipe modes or trapped cavity modes. A mode selective electronics system for all 3 options is currently under design. After building and testing it at FLASH, the performance of each option will be evaluated. Tests with the diode-based system will continue. Depending on financial issues, HOM-BPMs may also be built for the 3$^{rd}$ harmonics cryo-module at the European X-ray Free Electron Laser (XFEL), containing 8 cavities, as well as for a select number of 1.3 GHz modules. The outcome of this work may have a positive impact for other accelerators using SC RF technology.